\documentstyle[12pt]{article}
 \newcommand{\be}{\begin{eqnarray}}
 \newcommand{\ee}{\end{eqnarray}}
 \newcommand{\nee}{\nonumber\end{eqnarray}}

\begin{document}
\begin{titlepage}
\begin{flushright}
hep-ph/9907265\\
INRNE-TH 33/99\\
July 1999
\end{flushright}
\vfill
\begin{center}
{\Large\bf Semi-inclusive $\pi^\pm$ production -- tests for
independent fragmentation and for  polarized quark densities}
\\
\vspace{2cm}
{\large\bf Ekaterina Christova} \\
{\em Institute of Nuclear Research and Nuclear Energy, \\
 Boul. Tzarigradsko Chaussee 72, Sofia 1784, Bulgaria \\
 e-mail: echristo@inrne.bas.bg}\\

\vspace{1cm}
{\large\bf Elliot Leader} \\
{\em Deparment of Physics\\
BirBeck Colledge, University of London\\
Malet street., London WC1E 7HX, England\\
e-mail: e.leader@physics.bbk.ac.uk}
\end{center}
\vfill

\begin{abstract}
We show that measurements of semi-inclusive $\pi^\pm$ asymmetries
on $p$ and $n$ with polarized and
unpolarized target and beams allow, without any
knowledge of the  polarized parton densities 1) to test
independent fragmentation and SU(2) symmetry for the polarized sea, and 2) to
determine separately the polarized strange
 and  valence quarks densities, and the
ratio of the fragmentation functions $D_s^{\pi^++\pi^-}/D_u^{\pi^++\pi^-}$.
\end{abstract}
\end{titlepage}
\newpage
\setcounter{page}{1}

\section{Introduction}

Recently measurements of semi-inclusive processes with polarised
and unpolarised initial beams have been performed~\cite{EMC,factor,SMC}.
 New, more precise data are  expected soon. The goal
of these experiments is to determine better the parton
distribution functions,
 to distiguish between the quark and
antiquark contributions and to understand better the
fragmentation of quarks into hadrons. One of the basic
theoretical elements assumed in these studies is the
factorization of the process into quark production followed by
independent quark fragmentation into hadrons. We shall refer to
this as independent fragmentation.
Some  possibilities for a test of independent fragmentation
 have been considered~\cite{factor,de'Florian}.

In this paper we consider $\pi^++\pi^-$ semi-inclusive deep
inelastic lepton nucleon
 scattering with both polarized and unpolarized initial particles. We
 discuss different tests for independent fragmentation.
 Further, assuming it to be valid
 we formulate  tests for the polarized sea
and valence quark distribution functions, and for some general
symmetries among them. Apart from independent fragmentation
the tests formulated in this paper require  no other
assumptions. They
are general and independent of the parametrization of the
polarized quark densities and fragmentation functions. This
is achieved by finding  combinations of the cross sections on $p$ and $n$
for  semi-inclusive $(\pi^++\pi^-)$
production,  both for polarized and unpolarized beams, which
allow the use of
the data on inclusive  deep inelastic scattering (DIS)
 with polarized and unpolarized beam and target
and thus to formulate tests that involve only directly
measurable quantities. We work in the framework  of the
Parton Model in the current fragmentation region~\cite{Graudenz}, 
which implies
that we  consider  $\pi^{\pm}$ mesons produced mainly in the
forward direction.

\section{Testing independent fragmentation}

Consider the sum of the cross sections for producing $\pi^+$ and
$\pi^-$ in the semi-inclusive deep inelastic scattering
\be
l + N \rightarrow l' + \pi + X
\ee
in both the unpolarized case, and when beam and target are
longitudinally polarized.

We shall use the notation
\be
\tilde \sigma \equiv \frac{x(P+l)^2}{4\pi
\alpha^2}\left (\frac{2y^2}{1+(1-y)^2}\right ) \frac{d^3\sigma}{dx\,dy\,dz}
\ee
and
\be
\Delta \tilde \sigma \equiv \frac{x(P+l)^2}{4\pi
\alpha^2}\left (\frac{y}{2-y}\right ) \left
[\frac{d^3\sigma_{+-}}{dx\,dy\,dz} -
\frac{d^3\sigma_{++}}{dx\,dy\,dz}\right ]
\ee
where $P^\mu$ and $l^\mu$ are the nucleon and lepton four
momenta, and $\sigma_{\lambda\mu}$ refer to a lepton of helicity
$\lambda$ and a nucleon of helicity $\mu$. The variables $x$, $y$,
$z$ are the usual DIS kinematic variables~\cite{de'Florian}.

Assuming independent fragmentation, and using isospin  and charge
conjugation invariance for the fragmentation functions, i.e.
\be
D_u^{\pi^+ + \pi^-}\equiv D_u^{\pi^+} + D_u^{ \pi^-} = D_d^{\pi^+
+ \pi^-}
\ee
and
\be
D_s^{\pi^+ + \pi^-} = D_{\bar s}^{\pi^+ + \pi^-}
\ee
we obtain  for the polarized case in lowest order of QCD:
\be
&&\Delta \tilde\sigma_p^{\pi^++\pi^-} =
\frac{1}{9} \left\{\left [ 4(\Delta u +
\Delta\bar u) + (\Delta d +
\Delta\bar d) \right ]D_u^{\pi^++\pi^-} + 2\Delta s D_s^{\pi^++\pi^-}
\right\}\\
&&\Delta \tilde\sigma_n^{\pi^++\pi^-} =
 \frac{1 }{9} \left\{\left [4(\Delta d +
\Delta\bar d) + (\Delta u +
\Delta\bar u) \right ]D_u^{\pi^++\pi^-} + 2\Delta s D_s^{\pi^++\pi^-}\right\}.
\ee
We eliminate the $s$-quark contribution by the difference:
\be
\Delta \tilde\sigma_p^{\pi^++\pi^-} - \Delta \tilde\sigma_n^{\pi^++\pi^-}
=  \frac{1}{3} \{(\Delta u +
\Delta\bar u) - (\Delta d +\Delta\bar d) \}D_u^{\pi^++\pi^-}.
\ee
Analogously for the cross section of ($\pi^+ +\pi^-$) for the
unpolarized  case we have:
\be
&& \tilde\sigma_p^{\pi^++\pi^-} = \frac{1}{9} \left\{\left [4(  u +
 \bar u) + (  d +
 \bar d) \right ]D_u^{\pi^++\pi^-} + 2  s D_s^{\pi^++\pi^-}\right \}\\
&&  \tilde\sigma_n^{\pi^++\pi^-} = \frac{1}{9} \left\{\left [4(d +
 \bar d) + (  u +
 \bar u) \right ]D_u^{\pi^++\pi^-} + 2  s D_s^{\pi^++\pi^-}\right\}.
\ee
We again eliminate the $s$-quark contribution by the difference:
\be
\tilde  \sigma_p^{\pi^++\pi^-} -  \tilde \sigma_n^{\pi^++\pi^-}
=  \frac{1}{3} \{(  u +
 \bar u) - (  d + \bar d) \}D_u^{\pi^++\pi^-}.
\ee
 Then we consider the asymmetry $\Delta R_{np}^{\pi^++\pi^-}$:
\be
\Delta R_{np}^{\pi^++\pi^-}(x,z,Q^2)
&\equiv&
\frac{\Delta \tilde\sigma_p^{\pi^++\pi^-} -
\Delta \tilde\sigma_n^{\pi^++\pi^-}}
{\tilde\sigma_p^{\pi^++\pi^-} - \tilde\sigma_n^{\pi^++\pi^-}}=\\
&=&\frac{(\Delta u +\Delta\bar u) - (\Delta d +\Delta\bar d) }
{( u +\bar u) - (d +\bar d) }(x,Q^2).\label{deltaR}
\ee

The l.h.s of (\ref{deltaR}) is, in principle, a function of
 $x$, $z$ and $Q^2$, but as a consequence of independent
fragmentation, it
should be in practice a function only of  $x$ and $Q^2$.

Moreover, the r.h.s. of  (\ref{deltaR}) can be expressed,
 without any assumptions about the
sea quarks in terms of the measured quantities of inclusive
 DIS. Namely:
\be
g_1^p - g_1^n &=&\frac{1}{6}\left [(\Delta u +\Delta\bar u) -
 (\Delta d +\Delta\bar d)\right ]\label{g1}\\
F_1^p - F_1^n &=&\frac{1}{6} \left [( u +\bar u) - (d +\bar
d)\right ]\label{F1}.
\ee
We obtain
\be
\Delta R_{np}^{\pi^++\pi^-}(x,z,Q^2) =
\frac{(g_1^p-g_1^n)}{F_1^p-F_1^n}(x,Q^2).\label{deltaRnp}
\ee
Hence, a test of (\ref{deltaRnp}) would permit a test of
independent fragmentation  that does not require
 any knowledge of the parton distribution functions.
The advantage as compared to previous tests of factorization with
unpolarized~\cite{factor} and polarized beam and
targets~\cite{de'Florian} is
that the r.h.s. of (\ref{deltaR}) is expressed directly in terms
of  measurable quantities.

One can formulate also an integrated version of (\ref{deltaRnp}). 
 If we define
\be
\Delta N_{p,n}^{\pi^+ + \pi^-} &=& \int_{z_1}^{z_2}dz \int_0^1dx
\,\Delta \tilde\sigma_{p,n}^{\pi^+ + \pi^-},\\
 N_{p,n}^{\pi^+ + \pi^-} &= &\int_{z_1}^{z_2}dz \int_0^1dx
 \,\tilde\sigma_{p,n}^{\pi^+ + \pi^-},
\ee
then analogous to  (\ref{deltaRnp}) we obtain
\be
\frac{\Delta N_{p}^{\pi^+ + \pi^-}-\Delta N_{n}^{\pi^+ + \pi^-}}
{ N_{p}^{\pi^+ + \pi^-}-N_{n}^{\pi^+ + \pi^-}}=
\frac{g_A/g_V}{3S_G}\label{int}
\ee
independent of $z_1$ and $z_2$.
 In (\ref{int}) we have used the Bjorken sum rule:
\be
\int_0^1 dx \left [( \Delta u + \Delta \bar u ) - ( \Delta d +
\Delta \bar d )\right ]=\frac{g_A}{g_V},
\ee
and Gottfried sum rule:
\be
S_G =  \int_0^1\frac{F_2^p -F_2^n}{x} dx = \frac{1}{3} +
\frac{2}{3}\int_0^1(\bar u -\bar d ) dx.
\ee
The value of $g_A/g_V$ is determined with a very good precission
$g_A/g_V = 1.2573\pm 0.0028$,
and recent measurements of DIS give the following value for $S_G$~\cite{SG}:
\be
S_G = 0.235 \pm 0.026
\ee
In the following  we  assume independent fragmentation  and formulate some
methods of determining  the quark densities and fragmentation functions.

\section{Measurement of the polarized strange quark densities}

One of the main uncertainties, in determining the parton
distribution functions from the data on DIS scattering,
 is the contribution of the $s$-quarks.
Different assumptions have been made in fitting the data~\cite{Elliot}.
Here we shall show that if  we  consider the asymmetries on $p$ and $n$
independently, we
can single out the $\Delta s$-quark contribution.

We suppose that the proton and neutron asymmetries
\be
&&A_{1p}^+ \equiv A_{1p}^{\pi^+ + \pi^-}=
\frac{\Delta \tilde \sigma_p^{\pi^+ + \pi^-}}
{\tilde\sigma_p^{\pi^+ + \pi^-}} \label{A1p}\\
&&A_{1n}^+ \equiv A_{1n}^{\pi^+ + \pi^-} =
\frac{\Delta \tilde\sigma_n^{\pi^+ + \pi^-}}
{\tilde\sigma_n^{\pi^+ + \pi^-}} \label{A1n}
\ee
are known. Then, after some algebra from (\ref{g1}), (\ref{F1}),
(\ref{A1p}) and (\ref{A1n}) we obtain expressions for both
$\Delta s$ and the ratio $D_s^{\pi^+ + \pi^-}/D_u^{\pi^+ +
\pi^-}$ in terms of measured quantities only:
\be
\frac{\Delta s}{s} &=& \frac{ A_{1p}^+  A_{1n}^+ (F_1^n-F_1^p) +
g_1^p A_{1n}^+ - g_1^n A_{1p}^+}
{(g_1^p -g_1^n) - (A_{1p}^+ F_1^p -A_{1n}^+ F_1^n)}\label{Deltas}\\
\frac{D_s^{\pi^+ + \pi^-}}{D_u^{\pi^+ + \pi^-}} &=& 1 +
\frac{9\left [ (g_1^p -g_1^n) - (A_{1p}^+ F_1^p -A_{1n}^+
F_1^n)\right ]}{s(A_{1p}^+ - A_{1n}^+)}\label{Ds}
\ee
The consistency of using (\ref{Deltas}) and (\ref{Ds}) to
determine $\Delta s/s$ and  $D_s^{\pi^+ + \pi^-}/D_u^{\pi^+ +
\pi^-}$ can be checked experimentally, since in (\ref{Deltas}) the
l.h.s. should not depend upon $z$, and in (\ref{Ds}) the l.h.s.
should not depend upon $x$.


\section{SU(2) symmetry for the sea quarks?}

As it is well known,  electromagnetic inclusive DIS cannot
yield information on the sea quark (i.e. antiquark) densities,
for the simple reason that only the combinations $\Delta q +
\Delta\bar q$ occur in the expressions for the observables. For
various reasons it is sometimes useful to parametrise separately
the valence and  sea-quark densities when analysing inclusive
DIS. Usually, then, some simplifying assumption is made, e.g.
an  SU(2) or SU(3) symmetric polarized sea. Of course these
assumptions cannot be tested in inclusive DIS and it is thus of
interest to study the sea via semi-inclusive DIS.

We shall show how the data on the (${\pi^+ + \pi^-}$) and
$({\pi^+ - \pi^-}$) asymmetries provide a general test of the
SU(2) invariance of the polarized sea quarks. We shall {\it not} assume
SU(2) invariance for the unpolarized sea.

Assuming  SU(2) for the polarized sea, (\ref{deltaR}) equals
\be
\Delta R_{np}^{\pi^++\pi^-}(x,z,Q^2)
=\frac{\Delta u_V - \Delta d_V }{ (u + \bar u ) - (d + \bar d)}\,(x,Q^2)
\ee
On the other hand, by a slight rewriting of the results of
de'Florian et al.~\cite{de'Florian},  the  polarized
valence quark
densities can be obtained
 without any assumtions
 about the sea quarks from
\be
\frac{\Delta \tilde\sigma_p^{\pi^+-\pi^-} -
\Delta \tilde \sigma_n^{\pi^+-\pi^-}}
{\tilde\sigma_p^{\pi^+-\pi^-} - \tilde \sigma_n^{\pi^+-\pi^-}}=
\frac{\Delta u_V - \Delta d_V }{ u_V - d_V }.
\ee

It follows that if SU(2) holds for the polarized sea, then
\be
\frac{\Delta \tilde\sigma_p^{\pi^++\pi^-} - \Delta
\tilde\sigma_n^{\pi^++\pi^-}}{ \tilde\sigma_p^{\pi^++\pi^-} -
\tilde \sigma_n^{\pi^++\pi^-}} = \left [\frac{u_V-d_V}{(u + \bar u ) - (d +
\bar d)}\right ]\frac{\Delta \tilde\sigma_p^{\pi^+-\pi^-} - \Delta
\tilde\sigma_n^{\pi^+-\pi^-}}{ \tilde\sigma_p^{\pi^+-\pi^-} -
\tilde \sigma_n^{\pi^+-\pi^-}}.
\ee
Thus, without requiring any knowledge of the {\em polarized} densities
we can test whether the polarized sea is SU(2) symmetric.

\section{Measurements of the polarized valence quark densities}

For completeness we give finally the expressions relating the
polarized valence densities to the observables. We have:
\be
\frac{\Delta u_V }{ u_V }(x, Q^2)
=
\frac{4\Delta \tilde\sigma_p^{\pi^+-\pi^-} + \Delta \tilde
\sigma_n^{\pi^+-\pi^-}}
{4\tilde \sigma_p^{\pi^+-\pi^-} +
\tilde\sigma_n^{\pi^+-\pi^-}}\,(x,z,Q^2) \label{uV}
\ee
and
\be
\frac{\Delta d_V }{ d_V }(x, Q^2)=
\frac{\Delta \tilde\sigma_p^{\pi^+-\pi^-} +4 \Delta \tilde
\sigma_n^{\pi^+-\pi^-}}
{\tilde\sigma_p^{\pi^+-\pi^-} +4 \tilde\sigma_n^{\pi^+-\pi^-}}\,
(x,z,Q^2). \label{dV}
\ee
In order to improve statistics one may prefer to replace the
numerator and denominator in (\ref{uV}) and (\ref{dV}) by the
relevant cross-section differencies integrated over some range of
$z$.
\section{Conclusions}

It will be difficult to obtain accurate information on the
various sums and differencies of $\pi^\pm$ asymmetries in
semi-inclusive DIS with both proton and neutron targets.
Nevertheless  very important
information can be extracted. Without using any knowledge of the
parton densities one can i) test factorization of the production
and fragmentation of a quark, i.e. the concept of independent
fragmentation, ii) determine the polarized strange
quark densities $\Delta s$ and the ratio of the fragmentation
functions $D_s^{\pi^++\pi^-}/D_u^{\pi^++\pi^-}$, and iii) measure
the valence polarization densities $\Delta u_V/du_V$ and $\Delta
d_V/d_V$.  Finally, using information on the unpolarized $u$ and
$d$ quark densities yields a test of SU(2) symmetry of the
polarized sea.
\section{Acknowledgements}

The authors are grateful to the UK Royal Society for a
Collaborative Grant. E.C.'s work was  supported by the
Bulgarian National Science Foundation. E.L. is grateful for the
hospitality of the Department of Physics and Astronomy, the Vrije
University, Amsterdam, where part of this work was carried out,
supported by the Foundation for Fundamental Research on Matter
(FOM) and the Dutch Organization for Scientific Research (NWO).

\section{Note added in proof}

After submission of the paper our attention has been drawn to a paper
by L. Frankfurt et al. \cite{Frankfurt}, which addresses related issues.



\newpage
\begin{thebibliography}{99}
\bibitem{EMC} M. Arneodo et al. (EMC collaboration),
Z. Phys. {\bf C31} (1986) 1,
 Nucl. Phys. {\bf B321} (1989) 541
\bibitem{factor} J.J. Aubert et al. (EMC collaboration) Phys.
Lett. {\bf B114} (1982) 373, Z. Phys. {\bf C31} (1986) 175
\bibitem{SMC} B. Adela et al. (SMC), Phys. Lett. {\bf B369}
(1996) 93, Phys. Lett. {\bf B420} (1998) 180,\\
 W. Melnitchouk, J. Speth, A.W. Thomas, Phys. Lett. {\bf B435} (1998) 420,\\
  W. Melnitchouk, hep-ph 9906488
\bibitem{de'Florian} D. de Florian, L.N. Epele, H. Fanchiotti,
C.A. Garcia Canal, S. Joffily, R. Sassot,  Phys. Lett. {\bf B389}
(1996) 358
\bibitem{Graudenz} D. Graudenz , Nucl. Phys. {\bf B432} (1994) 351\\
D. de Florian, C.A. Garcia Canal, R. Sassot, Nucl. Phys. {\bf B470}
 (1996) 195\\
 D. de Florian, O.A.Sampayo, R. Sassot, Phys. Rev. {\bf D57}
(1998) 5803
\bibitem{SG} A. Doyle, talk given at ICHEP'98, Vancouver hep-ex/9812029
\bibitem{Elliot}
G. Altarelli, R.D. Ball, S. Forte, G. Ridolfi, Nucl. Phys.
{\bf B496} (1997) 337, Acta Phys. Polon. {\bf B29} (1998) 1145\\
E. Leader, A. Sidorov, D. Stamenov Phys. Rev.
{\bf D58} (1998) 114028, Int. J. Mod. Phys. {\bf A13} (1998) 5573\\
C. Bourrely, F. Buccella, O. Pisanti, P. Santorelli, J.
Soffer, Prog. Theor. Phys. {\bf 99} (1998) 1017
\bibitem{Frankfurt}  L. Frankfurt et al, Phys. Lett. {\bf B230}
(1989) 141
\end{thebibliography}
\end{document}